# External Factors that Affect the Photoplethysmography Waveforms

*Irene Pi[1], Isleen Pi[1] and Wei Wu[2*]*

[1] *Torrey Pines High School, San Diego, California,*
[2] *Dept. of Electronical and computer Engineering, USC, California*
* Email: wu.w@usc.edu

**Abstract**

Photoplethysmography (PPG) is a simple and inexpensive technology used in many smart devices to monitor cardiovascular health. The PPG sensors use LED lights to penetrate into the bloodstream to detect the different blood volume changes in the tissue through skin contact by sensing the amount of light that hits the sensor. Typically the data is displayed on a graph and it forms the pulse waveform. The information from the produced pulse waveform can be useful in calculating measurements that help monitor cardiovascular health, such as blood pressure. With many more people beginning to monitor their health status on their smart devices, it is extremely important that the PPG signal is accurate. Designing a simple experiment with standard lab equipment and commercial sensors, we wanted to find how external factors influence the results. In this study, it was found that external factors, touch force and temperature, can have a large impact on the resulting waveform so the effects of those factors need to be considered in order for the information to become more reliable.

## 1.0 Introduction & Background

Constant monitoring of blood pressure is important for many patients, especially those with a family history of hypertension. In recent years, cuffless blood pressure [1,5,10,13,18,20,23] has become a more popular method to monitor cardiovascular health and many of them use the optical technique, Photoplethysmography (PPG). The PPG technique can be found in many commercial health monitoring devices [7] because it is inexpensive and simple to use.

The PPG method [17] uses a sensor that measures the amount of light that hits the sensor. We used two methods to collect data from the PPG sensor. One way uses the green LED (530nm) light built on the sensor and the sensor measures the amount of light that is reflected back on the same side. The other way is to block the green light from the sensor and use an external red LED (650nm) light from the opposite side, where the sensor will eventually be measuring the amount of light that is transmitted across the finger.

The resulting changes in light intensity are displayed in a waveform and this is directly related to the changes in blood volume. The pulse waveform is a very important part of this method. Previous study identified the two parts of the wave, the larger first peak (systolic peak) and the smaller second peak (diastolic notch). The first half, including the large peak, represents the closing of the mitral and tricuspid valve while the second half, including the smaller peak, represents the closing of the aortic and pulmonary valves. After the closing of each valve, blood is pushed out of the heart and into the arteries, where it will send out a pulse of blood pressure. Older people have stiffer arteries so they are more likely to have high blood pressure while younger people have more flexible arteries. This trend can be demonstrated using the PPG [2,9,12]. In addition, PPG waveforms are also influenced by the whole cardiovascular system and surrounding tissues. Using the shape of the waveform produced by each round of tests, scientists can predict the blood pressure of the subject using pulse arrival time (PAT) and pulse transition time (PTT) [4,8,19].

While there are many benefits to the PPG method, there have also been a few instances where unexpected errors occurred due to the sensor's sensitivity [14]. Additionally, there are a variety of variables that may negatively influence the results [5]. PPG waveform is known to be sensitive to the location of sensor on the body and motion of subject [19,20]. The light source wavelength also affects the waveform due to the different penetration depth of the light in human tissue [16]. Taking this into account, we designed a study that tests how some other external factors such as touch force, temperature, and sensor configuration affect the resulting waveform [11,15,21,24]. Through our experiments, we hope to find the issues with the PPG method and potentially improve the accuracy of the signals produced.

## 2.0 Methods/Experiment Setup

We used commercial sensors and standard lab equipment to measure the PPG signals. The following are the equipment that we used:

- LeCory LC534AM 1GHz Digital Oscilloscope [25]
- PPG sensor built with Green LED (530nm) light and a photodiode as detector. (https://pulsesensor.com/)
- Red LED (650nm) light is used for the transmission, (MTPS7065MC from Digikey).
- Digital scale used for measuring the pressure of the finger (0.1 – 300 g force).
- DC Power supply (0-12 VDC)

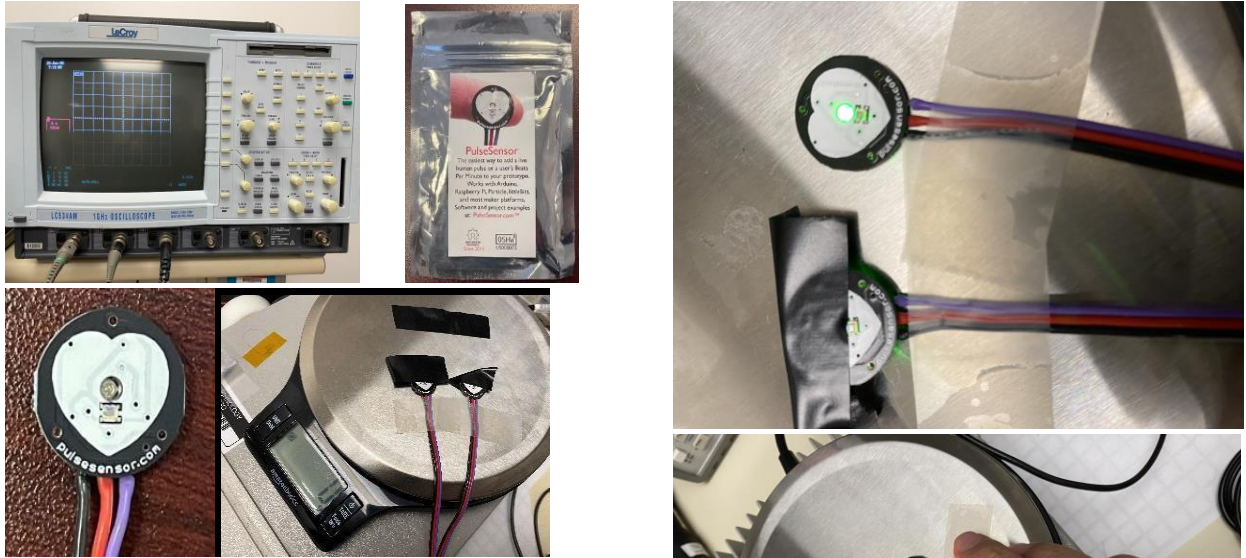

*Figure 1: Basic equipment used in the experiment*

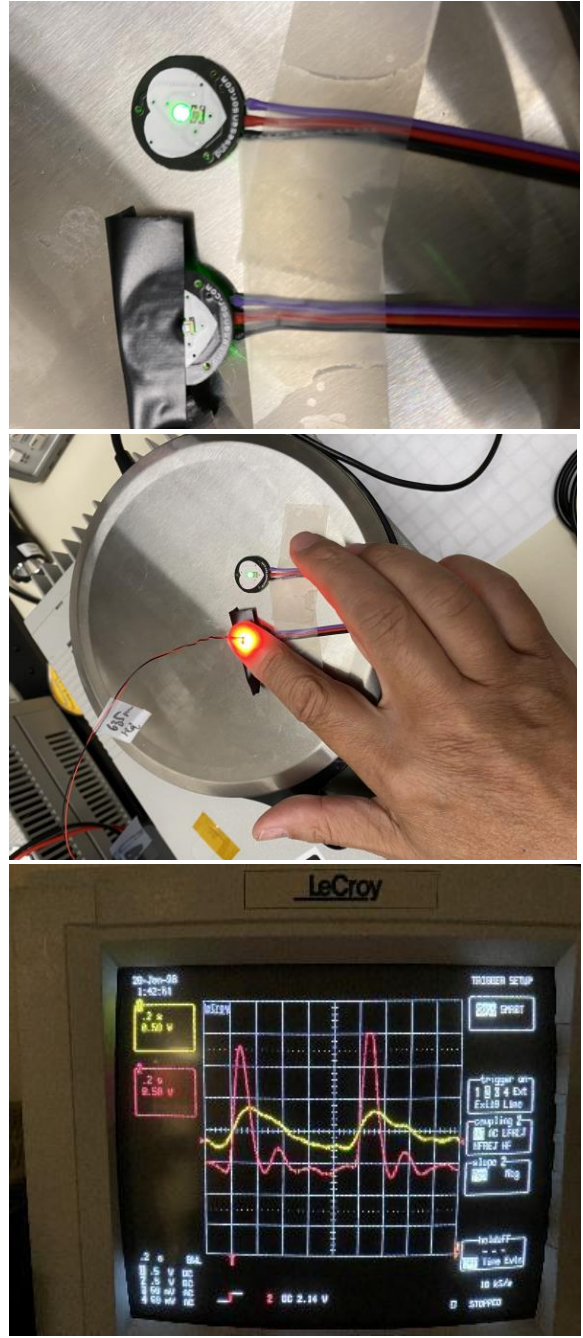

*Figure 2: (A) One PPG sensor with the green light exposed and another covered by black tape (B) Red light shines through finger and hits the sensor measuring the transmission signals (C) The measured signals shown on an oscilloscope*

As shown in Figure (2), we used a pulse sensor from pulsesensor.com to capture the waveforms and display it on the LeCory Oscilloscope. In the image, there are two pulse sensors displayed, one for reflection and another for transmission. We used a digital scale to measure the touch force of the finger on the sensor in grams. This will be used to check the changes to the waves based on touch force variation and to also ensure that the touch force is constant when other variables are measured.

As the amount of blood in the arteries change, the amount of light captured by the sensor also changes. The PPG sensor can monitor light intensity through two methods: transmission and reflection. When testing using the reflection method, the finger was placed on top of the sensor and the light from the sensor was used to determine the changes in light intensity, displaying it in the form of a wave. For the transmission method, light from the pulse sensor was blocked out and another red light LED was taped to the top of the finger to get a more accurate light intensity waveform.

We measured four subjects with age range from 17 to 55 years old. We measured the sensor output peak value to compare the different test conditions.

## 3.0 Results & Discussion

The first external variable that was explored was touch force. The measurement was done in reflective mode, where green LED (530nm) was used. Figure 3 shows the different pulse waveforms stacked on top of each other, each tested with the left index finger using different touch forces. The different forces were measured as the fingertip was placed on the sensor and scale at the 0.15N touch force (red), 0.44N touch force (blue), and the 0.78N touch force (green).

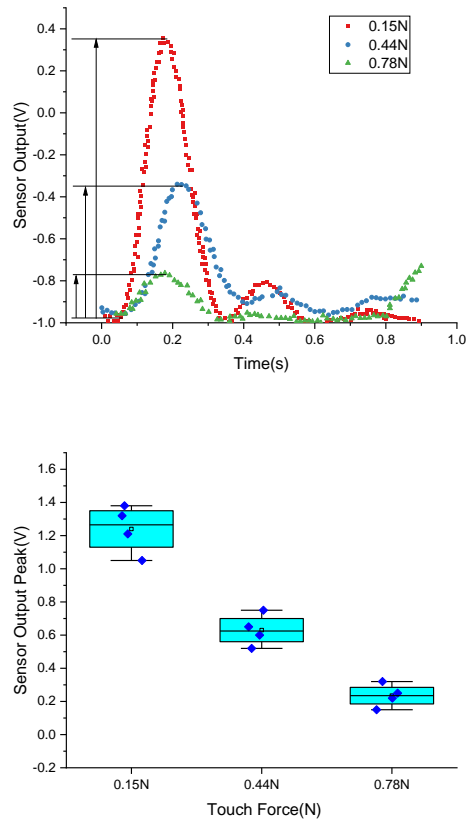

*Figure 3: A:Stacked PPG waveforms at different touch force pressures. B: Summary of the sensor responses (peak value) over different subjects.*

When the fingertip is gently placed on the sensor and the scale at around 0.15N touch force, the resulting waveform has the largest amplitude (red trace in Figure 3A) and this is also when there is lower blood flow resistance. As the test subject begins to press down on the sensor with a little more force to about 0.44N, the shape of the wave changes

(blue trace in Figure 3A), the amplitude decreases and the second peak becomes more apparent. When the finger touch force is increased to about 0.88N, the amplitude decreases even more (green trace in Figure 3A) and it becomes difficult to see the two part wave. We extracted signal peak amplitudes from four different test subjects. The results are show in Figure 3B. At a touch force that is too high, the blood flow is restricted and it may even be blocking the blood flow. At that point, it becomes difficult for the sensor to measure the changes in light intensity so the resulting waveform is missing some important information. This result is similar to some of the early studies, where different methods were used to study to touch effect. Since the shape of the wave widely varies when a touch force only changes a little, the data collected in this experiment support the idea that it is extremely difficult to accurately extract the necessary measurements to calculate the blood pressure.

The next factor that we tested was finger temperature. Figure 4A shows the two waveforms stacked on top of each other, one measured when the finger was kept at room temperature before the test (black trace) and one measured immediately after placing the finger in ice water for approximately one minute (red trace). The measurement was done in reflective mode, where green LED (530nm) light was used. When stacked on top of each other, it is apparent that the shape of the waveform changes with temperature. Again, we extracted the signal peak amplitude, and summarized four test subjects' results in the Figure 4B.

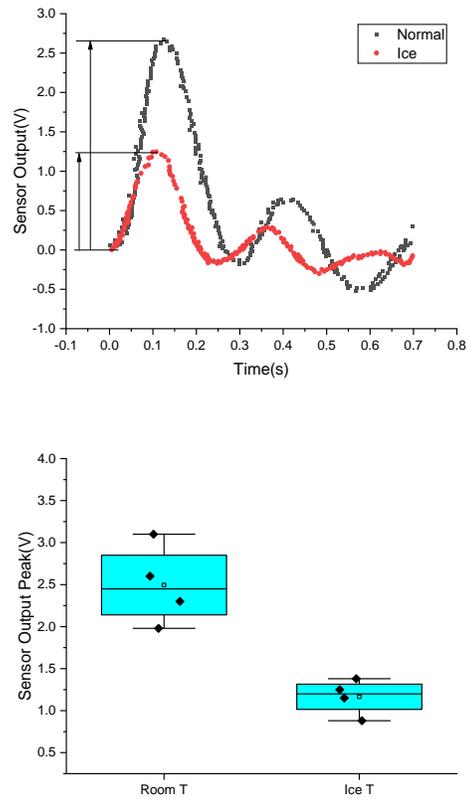

*Figure 4: A: Stacked PPG waveforms at different temperatures with a constant touch force of approximately 0.68N. B: Summary of four subjects' results.*

When the fingertip left at room temperature touches the sensor, the waveform that is formed shows a standard two-part wave. When the fingertip is pressed against the sensor after being placed in ice, the amplitude of the waveform significantly decreases and the entire wave occurs at a faster speed as the peaks are shifted towards the left. Since the amplitude of the wave decreases after the finger is placed in ice, this means that the measured blood flow also decreases. Due to the wide variations in the observed waves with a small change of temperature in the fingertip, it is difficult to obtain accurate PAT or PTT values that are necessary to calculate the blood pressure from the PPG.

After experimenting with the external variables, we conducted a comparison between transmission and reflection sensors. Figure 5A shows the two waveforms stacked on top of each other with a touch force of 0.44N, one measured using the transmission method (black trace) and one measured using the reflection method (red). Figure 5B shows the two waveforms stacked on top of each other with touch force of 0.88N, one measured using the transmission method (black trace) and one measured using the reflection method (red). Based on the results in the table, a visible difference in shape can be seen between the resulting waves.

At around 0.44N, the reflection (red) wave when compared to the transmission (blue) wave, has a larger amplitude but occurs at a similar rate. However, at around 0.88N, the observations are flipped. The reflection (red trace) waveform amplitude reduces dramatically while the transmission (black trace) waveform has a slightly larger amplitude. Since the reflection method only measures the surface blood vessels, it is more sensitive by the touch force and the waveform changes easily by touch force. By using the transmission method, the sensor takes the sum of all the light hit the sensor surface after passing through the entire finger. It is less sensitive to the touch force. At a lower force, there is a larger volume of blood in the finger, so less light is able to pass through the finger compared with when the touch force increases and there is less blood in the finger due to the pressure on finger. Thus, more of the light is transmitted through the finger, so the amplitude of the waveform increases. Figure 5B show the summary of waveform amplitude from four test subjects.

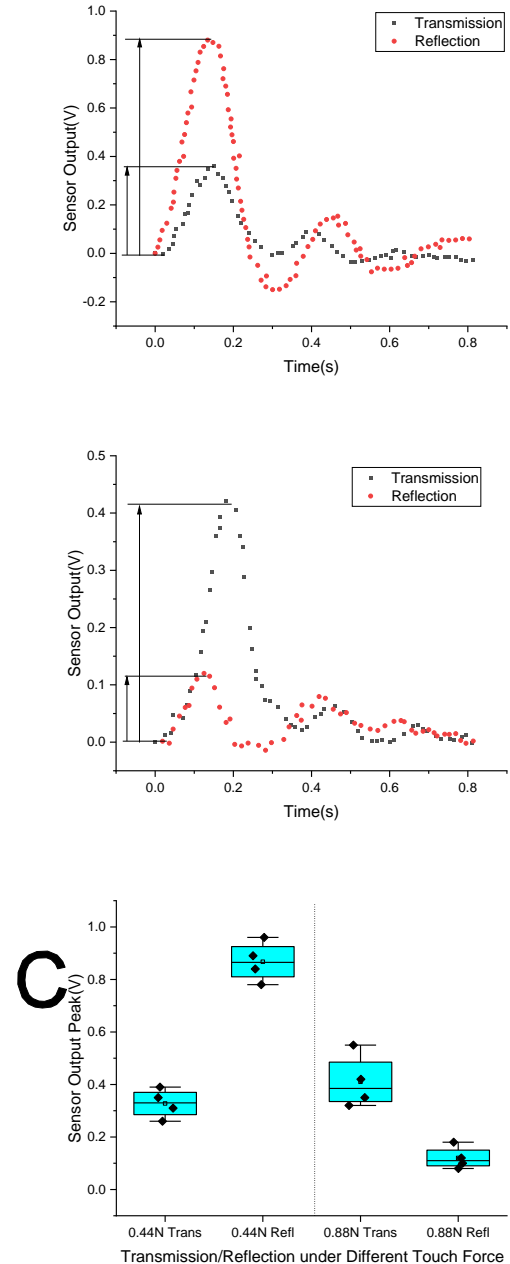

*Figure 5: A: Stacked PPG waveforms obtained using the transmission and reflection methods at 0.44N touch force. B: Waveforms obtained at*

*0.88N touch force. C: Summary of signal amplitude over four subjects*

Based on the findings, both transmission and reflection waves show the necessary points to calculate the blood pressure. Both models show touch pressure sensitivity, just with a different magnitude. While we used both reflection and transmission in the experiment, the resulting waves still had a few obvious corresponding points so scientists might be able to measure blood pressure based on only this wave. In general, transmission waveform seems a little less sensitive to touch pressure effect.

This experiment has some limitations due to time and resources. We could use more test subjects, and we could have built a better apparatus for more precise control of touch force. However, we believe that the results support the trend found in some of the other studies that there are many external factors PPG signal waveform. Any attempts to use complicated waveform analysis to deduct other physiological parameters need toconsider many correction/calibration factors.

## 4.0 Conclusion

Through this study, we have shown that the waveforms obtained from the PPG sensors are influenced by many factors including touch force and temperature. Both the temperature and touch force affects the blood vessels which also affects the measured blood pressure. While the waveform produced may be used to calculate the local blood pressure, it may be difficult to show the actual blood pressure of the test subject reliably. There are many other factors may affect the waveform. This may be concerning because the reflection PPG sensor is something that many smart devices use to calculate factors, like blood pressure, that determine cardiac health. Touch force and temperature are among many other factors that may negatively influence the end results, undermining the reliability of the measured numbers. Since it is difficult to make sure that the touch force and temperature is always constant, in order to make this method more accurate, more study is necessary.

Another conclusion that we can draw from our study is that the two different methods tested in this experiment, reflection and transmission, each produced different waveform shapes but with similar patterns. We have been able to show how the amplitude of each waveform changes with varying touch forces due to the setup differences between the two methods. Under the same conditions, both methods produced waveforms that had similar peaks, allowing experts to use this data to calculate blood pressure. Although both methods work, using transmission might actually be more accurate than the reflection method because the transmission mode data showed a little less sensitivity to touch force changes. This may be because the transmission signal takes the average of all blood vessels in the finger while reflection only takes measurements from the surface.

## 5.0 Declarations
**Funding**